\documentclass[article,12pt]{elsarticle}

\usepackage{amssymb}
\usepackage{amsthm}

\usepackage{color}
\usepackage{latexsym}
\usepackage{amsmath}
\usepackage{graphicx}
\journal{XXX}

\begin{document}

\begin{frontmatter}



\title{Addressing some critical aspects of the BepiColombo MORE relativity experiment }


\author[au1]{Giulia Schettino\corref{cor1}} 
\ead{g.schettino@ifac.cnr.it}
\author[au2]{Daniele Serra}
\ead{daniele.serra@dm.unipi.it}
\author[au2]{Giacomo Tommei} 
\ead{giacomo.tommei@unipi.it}
\author[au2]{Andrea Milani}
\ead{milani@dm.unipi.it}

\cortext[cor1]{Corresponding author}
\address[au1]{IFAC-CNR, Via Madonna del Piano 10, 50019 Sesto Fiorentino (FI), Italy}
\address[au2]{Dipartimento di Matematica, Largo B. Pontecorvo 5, 56127 Pisa, Italy}

\begin{abstract}
  The Mercury Orbiter radio Science Experiment (MORE) is one of the
  experiments on-board the ESA/JAXA BepiColombo mission to Mercury, to
  be launched in October 2018. Thanks to full on-board and on-ground
  instrumentation performing very precise tracking from the Earth,
  MORE will have the chance to determine with very high accuracy the
  Mercury-centric orbit of the spacecraft and the heliocentric orbit of
  Mercury. This will allow to undertake an accurate test of relativistic
  theories of gravitation (relativity experiment), which consists in
  improving the knowledge of some post-Newtonian and related
  parameters, whose value is predicted by General Relativity. This
  paper focuses on two critical aspects of the BepiColombo relativity
  experiment. First of all, we address the delicate issue of determining
  the orbits of Mercury and the Earth-Moon barycenter at the level of
  accuracy required by the purposes of the experiment and we discuss a
  strategy to cure the rank deficiencies that appear in the
  problem. Secondly, we introduce and discuss the role of the solar
  Lense-Thirring effect in the Mercury orbit determination problem 
  and in the relativistic parameters estimation.
\end{abstract}


\end{frontmatter}

\section{Introduction}
\label{intro}

BepiColombo is a space mission for the exploration of the planet
Mercury, jointly developed by the European Space Agency (ESA) and the
Japan Aerospace eXploration Agency (JAXA). The mission includes two
spacecraft: the ESA-led Mercury Planetary Orbiter (MPO), mainly
dedicated to the study of the surface and the internal composition of
the planet \cite{Benk}, and the JAXA-led Mercury Magnetospheric
Orbiter (MMO), designed for the study of the planetary magnetosphere
\cite{Mukai}. The two orbiters will be launched together in October
2018 on an Ariane 5 launch vehicle from Kourou and they will be
carried to Mercury by a common Mercury Transfer Module (MTM) using
solar-electric propulsion. The arrival at Mercury is foreseen for
December 2025, after 7.2 years of cruise. After the arrival, the
orbiters will be inserted in two different polar orbits: the MPO on a
$480\times 1500\,$km orbit with a period of 2.3 hours, while the MMO
on a $590\times 11639\,$km orbit. The nominal duration of the mission
in orbit is one year, with a possible one year extension.

The Mercury Orbiter Radio science Experiment (MORE) is one of the
experiments on-board the MPO spacecraft. The scientific goals of MORE
concern both fundamental physics and, specifically, the geodesy and
geophysics of Mercury. The radio science experiment will provide the
determination of the gravity field of Mercury and its rotational
state, in order to constrain the planet's internal
structure (\textit{gravimetry} and \textit{rotation
  experiments}). Details can be found, e.g., in
\cite{Mil_01,Sanchez,Iess_09,Cic_12,Cic_16,G_sait,G_17}. Moreover,
taking advantage from the fact that Mercury is the best-placed planet
to investigate the gravitational effects of the Sun, MORE will allow
an accurate test of relativistic theories of gravitation
(\textit{relativity experiment}; see, e.g.,
\cite{Mil_02,Mil_10,G_15,Schu,universe}). The global experiment
consists in a very precise orbit determination of both the MPO orbit
around Mercury and the orbits of Mercury and the Earth around the
Solar System Barycenter (SSB), performed by means of state-of-the-art
on-board and on-ground instrumentation \cite{Iess_01}. In particular,
the on-board transponder will collect the radio tracking observables
(range, range-rate) up to a goal accuracy (in Ka-band) of about
$\sigma_r=15\,$cm at 300 s for one-way range and
$\sigma_{\dot{r}}=1.5\times 10^{-4}\,$cm/s at 1000 s for one-way
range-rate \cite{Iess_01}. The radio observations will be further
supported by the on-board Italian Spring Accelerometer (ISA; see,
e.g., \cite{Iaf}). Thanks to the very accurate radio tracking,
together with the state vectors (position and velocity) of the
spacecraft, Mercury and the Earth, the experiment will be able to
determine, by means of a global non-linear least squares fit (see,
e.g., \cite{mil_gron}), the following quantities of general interest:
\begin{itemize}
\item coefficients of the expansion of Mercury gravity field in
  spherical harmonics with a signal-to-noise ratio better than 10 up
  to, at least, degree and order 25 and Love number $k_2$
  \cite{Kozai};
\item parameters defining the model of Mercury's rotation;
\item the post-Newtonian (PN) parameters $\gamma$, $\beta$, $\eta$,
  $\alpha_1$ and $\alpha_2$, which characterise the expansion of the
  space-time metric in the limit of slow-motion and weak field (see,
  e.g., \cite{Will_93}), together with some related parameters, as the
  oblateness of the Sun $J_{2\odot}$, the solar gravitational factor
  $\mu_{\odot}=GM_{\odot}$ (where $G$ is the gravitational constant
  and $M_{\odot}$ the mass of the Sun) and possibly its time
  derivative $\zeta=(1/\mu_{\odot})d\mu_{\odot}/dt$.
\end{itemize}

The aim of the present paper is to address two critical issues which
affect the BepiColombo relativity experiment and to introduce a
suitable strategy to handle these aspects. The first issue concerns
the determination of two PN parameters, the Eddington parameter
$\beta$ and the Nordtvedt parameter $\eta$. The criticality of
determining these parameters by ranging to a satellite around Mercury
has been already pointed out in the past (see, e.g., the discussion in
\cite{Mil_01} and \cite{ashby}). More recently, in \cite{DeMarchi} the
issue of how the lack of knowledge in the Solar System ephemerides can
affect, in particular, the determination of $\eta$ has been
discussed. Moreover, in \cite{GiuliaMA2016} the authors considered the
downgrading effect on the estimate of PN parameters due to
uncalibrated systematic effects in the radio observables and concluded
that these effects turn out to be particularly detrimental for the
determination of $\beta$ and $\eta$. Aside from these remarks, we observed
that the accuracy by which $\beta$ and $\eta$ can be determined turns
out to be very sensitive to changes in the epoch of the experiment in
orbit. Indeed, during the last years the simulations of the radio
science experiment in orbit have been performed assuming different
scenarios and epochs, due to the repeated postponement of the launch
date of the mission because of technical problems. As will be
described in the following, a deeper analysis reveals that the
observed sensitivity to the epoch of estimate is related to the rank
deficiencies found in solving simultaneously the Earth and Mercury
orbit determination problem, which affect in particular the estimate
of $\beta$ and $\eta$.

The second critical aspect we investigated concerns how the solar
Lense-Thirring (LT) effect affects the Mercury orbit determination
problem. The general relativistic LT effect on the orbit of Mercury
due to the Sun's angular momentum \cite{lense} is expected to be
relevant at the level of accuracy of our tests \cite{iorio2} and was
not included previously in our dynamical model (see a brief discussion
on this issue in \cite{universe}). Due to the resulting high
correlation between the Sun's angular momentum and its quadrupole
moment, we will discuss how the mismodelling deriving from neglecting
this effect can affect specifically the determination of $J_{2\odot}$.

The paper is organised as follows: in Sect.~\ref{sec:1} we describe
the mathematical background at the basis of our analysis, focusing on
the two highlighted critical issues. In Sect.~\ref{sec:2} we describe
how these issues can be handled in the framework of the orbit
determination software ORBIT14, developed by the Celestial Mechanics
group of the University of Pisa and we outline the simulation scenario
and assumptions. In Sect.~\ref{sec:3} we present the results of our
simulations and some sensitivity studies to strengthen the confidence
in our findings. Finally, in Sect.~\ref{concl} we draw some
conclusions and final remarks.

\section{Mathematical background}
\label{sec:1}

The challenging scientific goals of MORE can be fulfilled only by
performing a very accurate orbit determination of the spacecraft, of
Mercury and of the Earth-Moon barycenter (EMB)\footnote{The strategy
  adopted in our orbit determination code is to determine the EMB
  orbit instead of the Earth orbit.}.  Starting from the radio
observations, i.e. the distance (range) and the radial velocity
(range-rate) between the MORE on-board transponder and one or more
on-ground antennas, we perform the orbit determination together with
the parameters estimation by means of an iterative procedure based on
a classical non-linear least squares (LS) fit.

\subsection{The differential correction method}

Following, e.g., \cite{mil_gron} - Chap.~5, the non-linear LS fit aims
at determining a set of parameters $\mathbf{u}$ which minimises the
target function:
\begin{equation*}
Q(\mathbf{u})=\frac{1}{m}\boldsymbol{\xi}^T(\mathbf{u})W\boldsymbol{\xi}(\mathbf{u})\,,
\end{equation*}
where $m$ is the number of observations, $W$ is the matrix of the
observation weights and
$\boldsymbol{\xi}(\mathbf{u})=\mathcal{O}-\mathcal{C}(\mathbf{u})$ is
the vector of the residuals, namely the difference between the
observations $\mathcal{O}$ (i.e. the tracking data) and the
predictions $\mathcal{C}(\mathbf{u})$, resulting from the light-time
computation as a function of all the parameters $\mathbf{u}$ (see
\cite{lt} for all the details).

The procedure to compute the set of minimising parameters
$\mathbf{u}^\star$ is based on a modified Newton's method called {\it
  differential correction method}. Let us define the design
matrix $B$ and the normal matrix $C$:
\begin{equation*}
B=\frac{\partial \boldsymbol{\xi}}{\partial\mathbf{u}}(\mathbf{u})\,,\,\,\,\,\,\,C=B^TWB\,.
\end{equation*}
The stationary points of the target function are the solution of the
normal equation:
\begin{equation}
C\Delta\mathbf{u}^{\star}=-B^TW\boldsymbol{\xi}\,,
\label{norm_eq}
\end{equation}
where $\Delta\mathbf{u}^{\star}=\mathbf{u}^{\star}-\mathbf{u}$\,. The
method consists in applying iteratively the correction:
\begin{equation*}
\Delta\mathbf{u}=\mathbf{u}_{k+1}-\mathbf{u}_k=-C^{-1}B^TW\boldsymbol{\xi}
\end{equation*}
until, at least, one of the following conditions is met: $Q$ does not
change significantly between two consecutive iterations;
$\Delta\mathbf{u}$ becomes smaller than a given tolerance. In
particular, the inverse of the normal matrix, $\Gamma=C^{-1}$, can be
interpreted as the covariance matrix of the vector $\mathbf{u}^\star$
(see, e.g., \cite{mil_gron} - Chap.~3), carrying information on the
attainable accuracy of the estimated parameters.

The task of inverting the normal matrix $C$ can be made more difficult
by the presence of symmetries in the parameters space. A group $G$ of
transformations of such space is called group of \emph{exact symmetries} if,
for every $g\in G$, the residuals remain unchanged under the action of
$g$ on $\mathbf{u}$, namely:
\begin{equation*}
  \boldsymbol{\xi}(g[\mathbf{u}])=\boldsymbol{\xi}(\mathbf{u}).
\end{equation*}
It can be easily shown that if the latter holds, the normal matrix is
singular. In practical cases, the symmetry is usually
\emph{approximate}, that is there exists a small parameter $s$ such
that:
\begin{equation*}
\boldsymbol{\xi}(g[\mathbf{u}])=\boldsymbol{\xi}(\mathbf{u})+O(s^2)\,,
\end{equation*}
leading to a ill-conditioned normal matrix $C$, anyway yet
invertible. When this happens, solving for all the parameters involved
in the symmetry leads to a significant degradation of the results.
Possible solutions will be described in Sect.~\ref{subsec:1.1}.

\subsection{The dynamical model}

To achieve the scientific goals of MORE, both the Mercury-centric
dynamics of the probe and the heliocentric dynamics of Mercury and the
EMB need to be modelled to a high level of accuracy. On the one hand,
the MPO orbit around Mercury is expected to have a period of about 2.3
hours; on the other hand, the motion of Mercury around the Sun takes
place over 88 days. Thus, due to the completely different time scales,
we can handle separately the two dynamics. This means that, although
we are dealing with a unique set of measurements, we can conceptually
separate between gravimetry-rotation experiments on one side, mainly
based on range-rate observations, and the relativity experiment on the
other, performed ultimately with range measurements. Comparing the
goal accuracies for range and range-rate, scaled over the same
integration time according to Gaussian statistics, we indeed find that
$\sigma_r/\sigma_{\dot{r}}\sim 10^5\,$s. As a result, range
measurements are more accurate when observing phenomena with
periodicity longer than $10^5\,$s, like relativistic phenomena, whose
effects become significant over months or years. On the contrary,
since the gravity and rotational state of Mercury show variability
over time scales of the order of hours or days, the determination of
the related parameters is mainly based on range-rate observations.

All the details on the Mercury-centric dynamical model of the MPO
orbiter can be found in \cite{Cic_16} and \cite{universe}, including
the effects due to the gravity field of the planet up to degree and
order 25, the tidal effects of the Sun on Mercury (Love potential;
see, e.g., \cite{Kozai}), a semi-empirical model for the planet's
rotation (see \cite{Cic_12}), the third-body perturbations from the
other planets, the solar non-gravitational perturbations, like the solar
radiation pressure, and some non-negligible relativistic effects (see,
e.g., \cite{Moyer} and \cite{Cic_16}). In the following we will focus
on the relativity experiment, hence on the heliocentric dynamics of
Mercury and the EMB. In the slow-motion, weak-field limit, known as
Post-Newtonian (PN) approximation, the space-time metric can be
written as an expansion about the Minkowski metric in terms of
dimensionless gravitational potentials. In the parametrised PN
formalism, each potential term in the metric is characterised by a
specific parameter, which measures a general property of the metric
(see, e.g., \cite{will_LR}). Each PN parameter assumes a well defined
value (0 or 1) in General Relativity (GR). The effect of each term on
the motion can be isolated, therefore the value of the associated PN
parameter can be constrained within some accuracy threshold, testing
any agreement (or not) with GR.  The PN parameters that will be
estimated are the Eddington parameters $\beta$ and $\gamma$
($\beta=\gamma=1$ in GR), the Nordtvedt parameter $\eta$ \cite{nordt}
($\eta=0$ in GR) and the preferred frame effects parameters $\alpha_1$
and $\alpha_2$ ($\alpha_1=\alpha_2=0$ in GR). Moreover, we include in
the solve-for list a few additional parameters, whose effect on the
orbital motion can be comparable with that induced by some PN
parameters \cite{Mil_02}: the oblateness factor of the Sun
$J_{2\odot}$, the gravitational parameter of the Sun $\mu_\odot$ and
its time derivative $\zeta=(1/\mu_\odot)\, d\mu_\odot/dt$.

The modification of the space-time metric due to a single PN
parameter affects both the propagation of the tracking signal and the
equations of motion. As regards the observables, they must be computed
in a coherent relativistic background. This implies to account for the
curvature of the space-time metric along the propagation of radio
signals (Shapiro effect \cite{shap}) and for the proper times of
different events, as the transmission and reception times of the
signals. All the details concerning the relativistic computation of
the observables can be found in \cite{lt}.
A relativistic model for the motion of Mercury is necessary in order
to accurately determine its orbit and, hence, constrain the PN and
related parameters. The complete description of the relativistic
setting can be found in \cite{Mil_02,universe}.

\subsection{Determination of Mercury and EMB orbits}
\label{subsec:1.1}

\begin{figure}
  \includegraphics[width=0.50\textwidth]{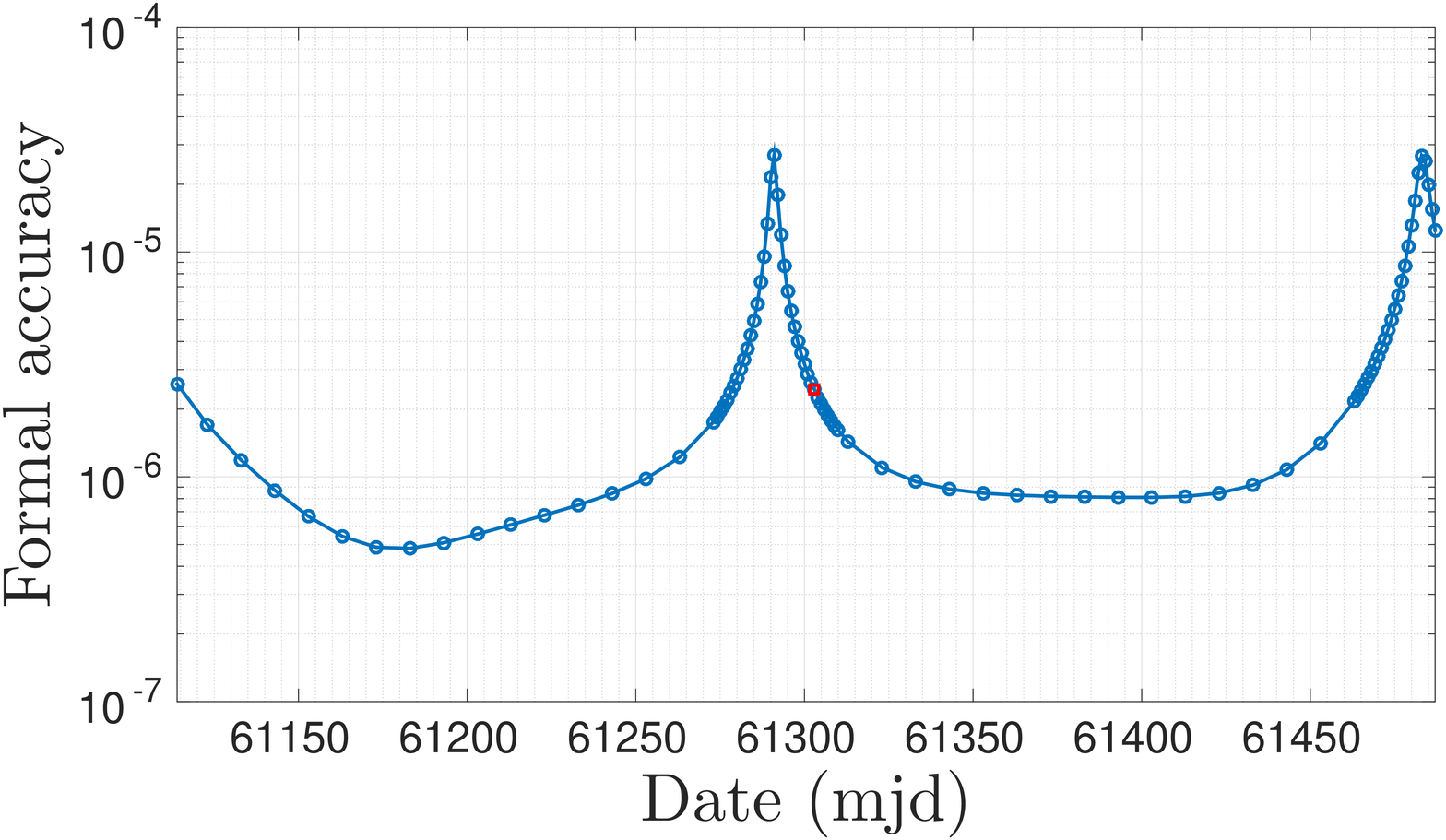}
  \includegraphics[width=0.50\textwidth]{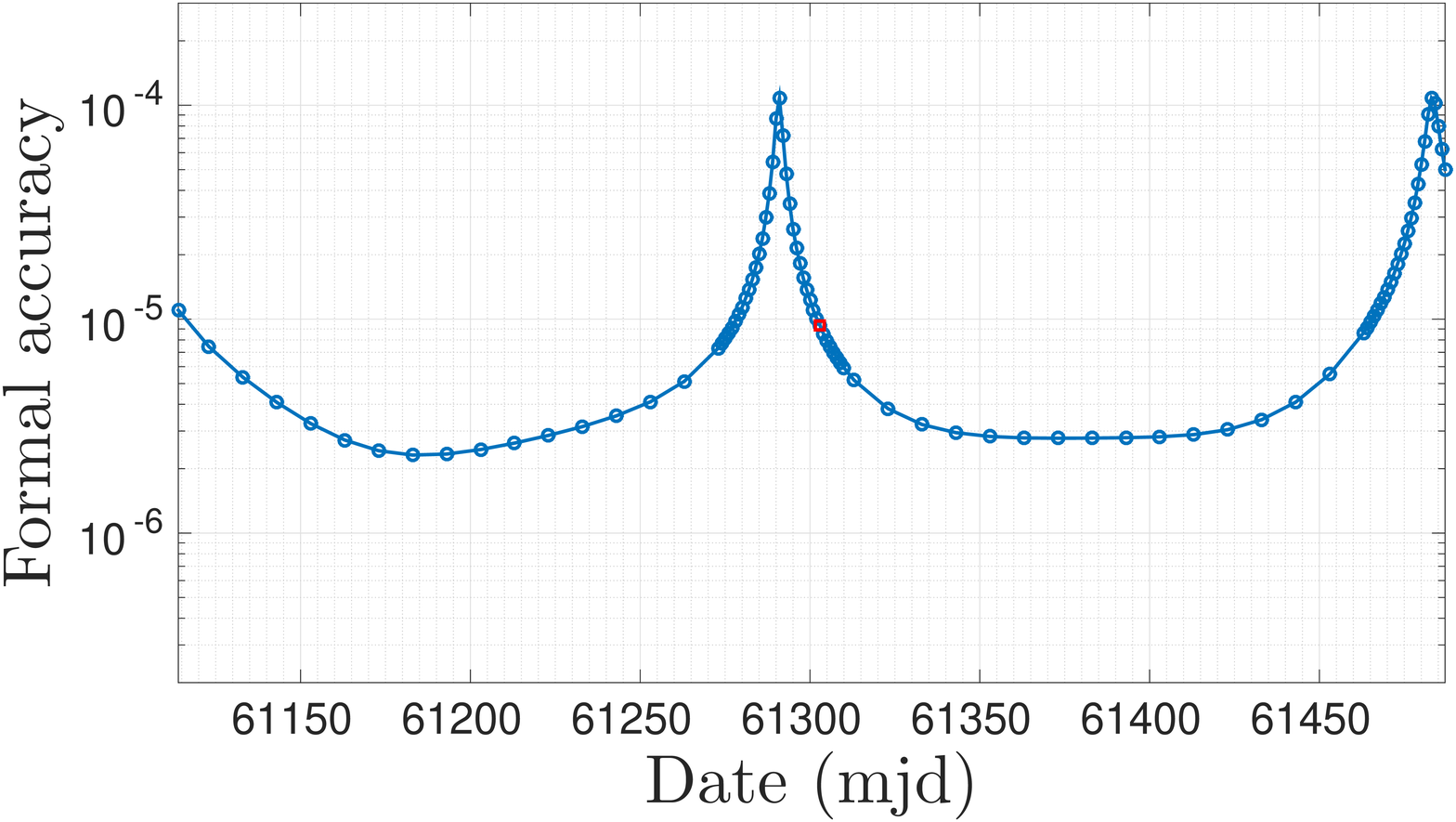}
  \caption{Formal accuracy of $\beta$ (left) and $\eta$ (right) as a
    function of the epoch of the estimate (in MJD) over the mission
    time span. In red the value of the accuracy for the estimate at
    central epoch.}
  \label{eta_NO}
\end{figure}

As already pointed out, the relativity experiment is based on a very
accurate determination of the heliocentric orbits of Mercury and
the EMB, that is we estimate the corresponding state vectors (position
and velocity) w.r.t. the SSB at a given reference epoch. A natural
choice is to determine the state vectors at the central epoch of the
orbital mission, whose duration is supposed to be one year. In
this way the propagation of the orbits is performed backward for the
first six months of the mission and forward for the remaining six
months, thus minimising the numerical errors due to propagation. Of
course, the determination of the PN and related parameters should not
depend significantly from the choice of the epoch of the estimate. To
verify this point, in Figure \ref{eta_NO} we have shown the behaviour
of the accuracy of $\beta$ (left) and $\eta$ (right), obtained from
the diagonal terms of the covariance matrix, as a function of the
epoch of the estimate, from the beginning of the orbital mission
(Modified Julian Date (MJD) 61114, corresponding to 15 March 2026) to
the end (MJD 61487, corresponding to 23 March 2027).
The value of the formal accuracy at the central epoch (MJD 61303,
corresponding to 20 September 2026) is highlighted in red. It is clear
that there is a strong dependency of the achievable accuracy on the
epoch of the estimate. If the planetary orbits are determined at MJD
61183 (23 May 2026), the accuracy of $\eta$ turns out to be
$\sigma(\eta)\simeq 2.3\times 10^{-6}$, whereas estimating at MJD
61291 (8 September 2026) results in $\sigma(\eta)\simeq 1.1\times
10^{-4}$, almost two orders of magnitude larger. On the contrary, the
uncertainty of the other PN parameters showed very little variability
with the epoch of the estimate.

Such behaviour indicates the presence of some weak directions in orbit
determination, possibly connected to the strategy adopted until now
for the MORE Relativity Experiment: we determine only 8 out of 12
components of the initial conditions of Mercury and EMB. This
assumption, first introduced in \cite{Mil_02}, is a solution to the
presence of an approximate rank deficiency of order 4, arising when we
try to determine the orbits of Mercury and the Earth (or, similarly,
the EMB as in our problem) w.r.t. the Sun only by means of relative
observations. Indeed, if there were only the Sun, Mercury and the
Earth, and the Sun was perfectly spherical ($J_{2\odot}=0$), there
would be an exact symmetry of order 3 represented by the rotation
group $SO(3)$ applied to the state vectors of Mercury and the
Earth. Because of the coupling with the other planets and due to the
non-zero oblateness of the Sun, the symmetry is broken but only by a
small amount, of the order of the relative size of the perturbations
of the other planets on the orbits of Mercury and the Earth and of the
order of $J_{2\odot}$.

Moreover, there is another approximate symmetry for scaling. The
symmetry would be exact if there were only the Sun, Mercury and the
Earth: if we change all the lengths involved in the problem by a
factor $\lambda$, all the masses by a factor $\mu$ and all the times
by a factor $\tau$, with the factors related by $\lambda^3=\tau^2\mu$
(Kepler's third law), then the equation of motion of the gravitational
3-body problem would remain unchanged. Since we can assume
\footnote{There are accurate definitions of the time scales based upon
  atomic clocks.} that $\tau=1$, the symmetry for scaling involves the
state vectors of Mercury and the Earth (i.e. the ``lengths'' involved
in the problem) and the gravitational mass of the Sun, which is among
the solve-for parameters. The symmetry for scaling can also be
expressed by the well known fact that it is not possible to solve
simultaneously for the mass of the Sun and the value of the
astronomical unit. Since the state vectors of the other planets,
perturbing the orbits of Mercury and the Earth, are given by the
planetary ephemerides and thus they cannot be rescaled, the symmetry
is broken but, again, only by a small amount. In conclusion, an approximate 
rank deficiency of order 4 occurs in the orbit determination problem we 
want to solve. Solving for all the 12 components of the
initial conditions and the mass of the Sun would result in
considerable loss of accuracy for all the parameters of the relativity
experiment, as will be quantified in Sect.~\ref{sec:3}.

The only solution in case of rank deficiency is to change the
problem. When no additional observations breaking the symmetry are
available, a convenient solution is to remove some parameters from the
solve-for list. Starting from $N$ parameters to be solved, in case of
a rank deficiency of order $d$, we can select a new set of $N-d$
parameters to be solved, in such a way that the new normal matrix
$\bar C$, with dimensions $(N-d)\times (N-d)$ instead of $N\times N$,
has rank $N-d$. The remaining $d$ parameters can be set at some
nominal value (\textit{consider parameters}). This solution has been
applied up to now in the MORE relativity experiment (see, e.g.,
\cite{universe}): the three position components and the out-of-plane
velocity component of the EMB orbit, for a total of 4 parameters, have
been removed from the solve-for list, curing in this way the rank
deficiency of order 4.

Another option can be investigated: the use of a priori
observations. When some information on one or more of the parameters
involved in the symmetry is already available -- for instance from
previous experiments -- it can be taken into account in our experiment
and could lead to an improvement of the results. In this case the
search for the minimum of the target function is restricted to the
vector of parameters fulfilling a set of a priori equations. In
practice, we add to the observations a set of a priori constraints,
$\mathbf{u}=\mathbf{u}^P$, on the value of the parameters, with given
a priori standard deviation $\sigma_i$ ($i=1,..N$) on each constraint
$u_i=u_i^P$. This is equivalent to add to the normal equation in
Eq.~(\ref{norm_eq}) an a priori normal equation of the form:
\begin{equation*}
C^P\mathbf{u}=C^P\mathbf{u}^P\,,
\end{equation*} 
with $C^P=\mbox{diag}[\sigma_i^{-1}]$. In this way, an ``a priori
penalty'' is added to the target function:
\begin{equation*}
Q(\mathbf{u})=\frac{1}{N+m}[(\mathbf{u}-\mathbf{u}^P)^TC^P(\mathbf{u}-\mathbf{u}^P)+\boldsymbol{\xi}^T(\mathbf{u})W\boldsymbol{\xi}(\mathbf{u})]
\end{equation*}
and the complete normal equation becomes:
\begin{equation*}
(C^P+C)\Delta\mathbf{u}=-B^TW\boldsymbol{\xi}+C^P(\mathbf{u}^P-\mathbf{u}_k)\,.
\end{equation*} 
If the a priori uncertainties $\sigma_i$ are small enough, the new
normal matrix $\bar{C}=C^P+C$ has rank $N$ and the complete orbit
determination problem can be solved.

In our problem, the a priori information is represented by four
constraint equations which inhibit the symmetry for rotation and
scaling, to be added to the LS fit as a priori observations. A
complete description of the form that the constraint equations assume
will be given in Sect.~\ref{subsec:2.1}.

\subsection{The solar Lense-Thirring effect}
\label{subsec:1.2}

In \cite{universe} we pointed out that the Lense-Thirring (LT) effect
on the orbit of Mercury due to the angular momentum of the Sun has
been neglected, in order to simplify the development and
implementation of the dynamical model. In fact the solar LT effect is
expected to be relevant at the level of accuracy of our tests
\cite{iorio2}. As it will be clear in Sect.~\ref{subsec:3.2}, the
mismodelling resulting from neglecting this effect affects
specifically the determination of the oblateness of the Sun,
$J_{2\odot}$.

\begin{figure}
  \includegraphics[width=\textwidth]{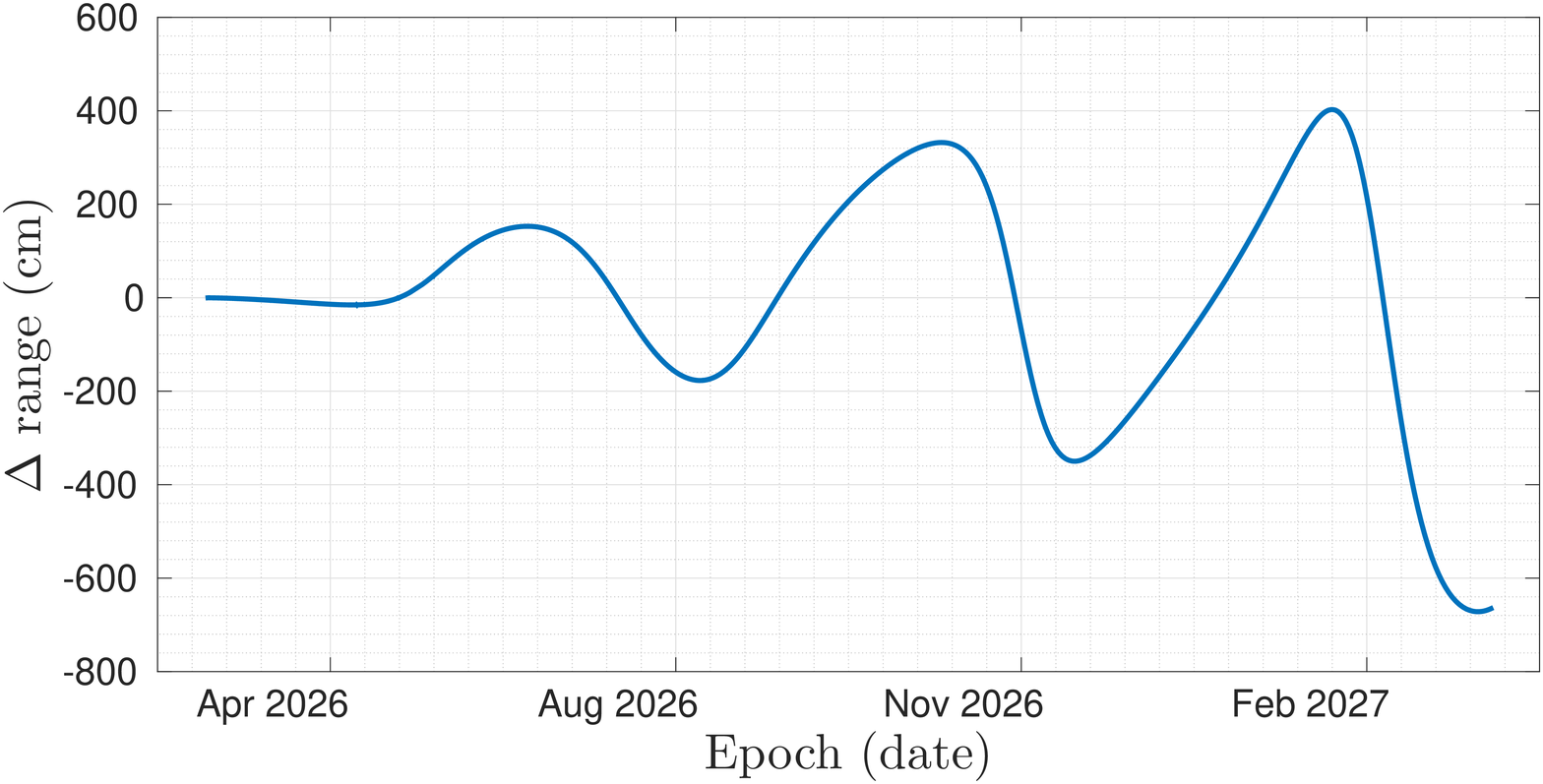}
  \caption{Difference (in cm) of simulated spacecraft range with and without
    solar LT perturbation in the dynamical model, over one-year
    mission time span.}
  \label{delta_range}
\end{figure}

More specifically, the general relativistic LT effect induces a
precession of the argument of the pericenter of Mercury in the gravity
field of the Sun at the level of
$\dot{\omega}_{\mathrm{LT}}=-2\,$milliarcsec/century, according to GR
\cite{iorio3}. We modelled the effect as an additional perturbative
acceleration in the heliocentric equation of motion of Mercury (see,
e.g. \cite{Moyer}):
 \begin{equation}
\mathbf{a}_{\mathrm{LT}}=\frac{(1+\gamma)\,GS_{\odot}}{c^2\,r^3}\left[-\hat{\mathbf s} \times \dot{\mathbf r}+ 3\,\frac{(\hat{\mathbf s} \cdot \mathbf r)\,(\mathbf r \times \dot{\mathbf r})}{r^2}  \right]\,,
\label{acc_LT}
\end{equation} 
where $\mathbf{S}_{\odot}=S_{\odot}\hat {\mathbf{s}}$ is the angular
momentum of the Sun ($\hat{\mathbf{s}}$ is assumed along the rotation
axis of the Sun). To assess the role of the solar LT in the dynamics,
in Figure \ref{delta_range} we plot the effect of the solar LT
acceleration, given by Eq.~(\ref{acc_LT}), on the simulated range of
the orbiter. In other words, this is the difference between simulated
range with and without LT effect over the one-year mission time
span. As it can be seen, the mismodelling due to the lack of the solar
LT perturbation in the dynamical model can be as high as some
meters. This result is in very good agreement with Fig.~1 in
\cite{iorio2}, which shows the numerically integrated EMB-Mercury
ranges with and without the perturbation due to the solar
Lense-Thirring field over two years in the ICRF/J2000.0 reference
frame, with the mean equinox of the reference epoch and the reference
$x-y$ plane rotated from the mean ecliptic of the epoch to the Sun’s
equator, centered at the SSB.

\section{The ORBIT14 software}
\label{sec:2}

Since 2007, the Celestial Mechanics Group of the University of Pisa
has developed\footnote{under an Italian Space Agency commission.}  a
complete software, ORBIT14, dedicated to the BepiColombo and Juno
radio science experiments \cite{tommei_J,serra}, which is now ready
for use. All the code is written in Fortran90. The software includes a
data simulator, which generates the simulated observables and the
nominal value for the orbital elements of the Mercury-centric orbit of
the MPO and the heliocentric orbits of Mercury and the EMB, and the
differential corrector, which is the core of the code, solving for the
parameters of interest by a global non-linear LS fit, within a
constrained multi-arc strategy \cite{Alessi}.  The general structure
of the software is described, e.g., in \cite{universe}.

\subsection{Handling the a priori constraints}
\label{subsec:2.1}

The equations needed to a priori constrain the LS solution are given as an
input to the differential corrector. In general, the $n$-th constraint
has the expression: $f_n(\mathbf{u})=0$. ORBIT14 has been designed to
handle only linear constraints. Thus, the equation for the $n$-th
constraint, involving $d$ parameters to be determined, reads:
\begin{equation*}
f_n(\mathbf{u})=\sum_{i=1}^{d}a_i(x_i-\theta_i)=N(0,\mbox{diag}[\sigma_i])\,,
\end{equation*}
where $\sigma_i$ are the weights associated to each parameter involved
in the constraint, assuming a Gaussian distribution with zero
mean. Following the notation of Sect.~\ref{subsec:1.1}, its
contribution to the normal matrix is given by:
\begin{equation*}
C_n^{P}=\left(\frac{\partial f_n}{\partial \mathbf{u}}\right)^T\,W\,\frac{\partial f_n}{\partial \mathbf{u}}\,,
\end{equation*}
and to the right hand side of the equations of motion by:
\begin{equation*}
D_n^{P}=\left(\frac{\partial f_n}{\partial \mathbf{u}}\right)^T\,W\,f_n\,,
\end{equation*}
where $W=\mbox{diag}[\sigma_i^{-2}]$

In order to write the linear constraint equations of our orbit determination
problem, let us introduce the following notation for the
components of the state vectors of Mercury and the EMB:
\begin{itemize}
\item $\bf{M}$, $\dot{\bf{M}}$: position and velocity of Mercury at
  the reference epoch from ephemerides (nominal values); $\bf{m}$,
  $\dot{\bf{m}}$: estimated position and velocity of Mercury;
  $\Delta\bf{M}=\bf{M}-\bf{m}$,
  $\Delta\dot{\bf{M}}=\dot{\bf{M}}-\dot{\bf{m}}$: deviation between
  ephemerides and estimate.
\item $\bf{E}$, $\dot{\bf{E}}$: position and velocity of EMB at the
  reference epoch from ephemerides; $\bf{e}$, $\dot{\bf{e}}$:
  estimated position and velocity of EMB;
  $\Delta\bf{E}=\bf{E}-\bf{e}$,
  $\Delta\dot{\bf{E}}=\dot{\bf{E}}-\dot{\bf{e}}$: deviation between
  ephemerides and estimate.
\end{itemize}
 
\paragraph{Symmetry for rotations.}

The symmetry for rotation is described by a three-parameter group,
whose generators are for example the rotations around three orthogonal
axis $(x,y,z)$ of the reference frame used for orbit propagation. The
constraint equation which inhibits an infinitesimal rotation by an
angle $s$ around the $x$-axis has the expression:
\begin{equation}
\begin{split}
& \left. \frac{\Delta\mathbf{M}}{|\mathbf{m}|}\cdot\frac{\partial (R_{s,\hat{x}}\hat{\mathbf{M}})}{\partial s}\right\arrowvert_{s=0}+\left. \frac{\Delta\mathbf{E}}{|\mathbf{e}|}\cdot\frac{\partial (R_{s,\hat{x}}\hat{\mathbf{E}})}{\partial s}\right\arrowvert_{s=0}+\left. \frac{\Delta\dot{\mathbf{M}}}{|\dot{\mathbf{m}}|}\cdot\frac{\partial (R_{s,\hat{x}}\hat{\dot{\mathbf{M}}})}{\partial s}\right\arrowvert_{s=0}+ \\
& + \left. \frac{\Delta\dot{\mathbf{E}}}{|\dot{\mathbf{e}}|}\cdot\frac{\partial (R_{s,\hat{x}}\hat{\dot{\mathbf{E}}})}{\partial s}\right\arrowvert_{s=0}=N(\mbox{diag}[\sigma_i],0)\,,
\end{split}
\label{eq_rot}
\end{equation}
where $\sigma_i$ are the weights for the state vectors components, $N$
represents a Gaussian distribution with zero mean, $R_{s,\hat{x}}$
is the rotation matrix by an angle $s$ around the $x$-axis:
\begin{equation*}
R_{s,\hat{x}}=
\begin{pmatrix}
1 & 0 & 0 \\
0 & \cos s & -\sin s \\
0 & \sin s & \cos s \\
\end{pmatrix}
\end{equation*}
and $(\partial R_{s}/\partial s) \arrowvert_{s=0}$ is a
generator of the Lie algebra of the rotations $SO(3)$.  Two similar
equations hold for the rotations by an angle $s$ around the $y$ and
$z$ axes.

\paragraph{Symmetry for scaling.}

To find the equation to constrain for scaling, we can start from the
simple planar two-body problem of a planet around the Sun, with the
non-linear dependency of the mean motion $n$ upon the semi-major axis
$a$, in the hypothesis of circular motion:
\begin{equation}
\frac{da}{dt}=0,\,\,\,\,\,\frac{d\lambda}{dt}=n(a)=\frac{k}{a^{3/2}}\,,
\end{equation} 
where $k^2=GM_{\odot}=\mu_{\odot}$, with solution given by:
\begin{equation}
a(t)=a_0,\,\,\,\,\,\lambda (t)=\frac{k}{a_0^{3/2}}t+\lambda_0\,.
\end{equation}
This problem has a symmetry with multiplicative parameter
$w\in \mathbb{R}^+$:
\begin{equation}
k \mapsto w^3 k,\,\,\,\,\, a_0\mapsto w^2a_0\,,
\end{equation}
leaving $n=k/a^{3/2}$ invariant. The symmetry can be represented
by means of an additive parameter $s$ by setting $w=e^{s}$. The derivative of the
symmetry group action with respect to $s$ is:
\begin{equation}
\frac{da_0}{ds}=2w^2a_0,\,\,\,\,\,\frac{dk}{ds}=3w^3k\,.
\end{equation}
Finally, the constraint takes the form:
\begin{equation}
-3\,\frac{\Delta a}{a_0}\,\left.\frac{da_0}{ds}\right\arrowvert_{s=0}+2\,\frac{\Delta k}{k_0}\,\left.\frac{dk}{ds}\right\arrowvert_{s=0}=0\,.
\end{equation}

In our fit, we estimate the parameter $\mu_{\odot}$, that is
$k^{1/2}$. Since we need to deal with linear constraints, we can
linearize the problem by expanding the non-linear equation up to the
first order around the nominal value.  In this way, the final
expression adopted for the scaling constraint reads:
\begin{equation}
\sum_{j=1}^3\left [ \frac{\Delta M_j}{|\mathbf{M}|}\,M_j+\frac{\Delta \dot{M}_j}{|\dot{\mathbf{M}}|}\,\dot{M}_j+\frac{\Delta E_j}{|\mathbf{E}|}\,E_j+\frac{\Delta \dot{E}_j}{|\dot{\mathbf{E}}|}\,\dot{E}_j\right ]+3\Delta\mu_{\odot}=N(\mbox{diag}[\sigma_i],0)\,,
\label{eq_scal}
\end{equation}
where $j=1,2,3$ refers to the three orthogonal directions $x,y,z$.

\paragraph{Setting the weights $\sigma_i$.}
Together with the constraint equations given in input to the
differential corrector, it is necessary to provide also the a priori
standard deviations $\sigma_i$ by which the involved parameters are
constrained. The strength of the weights $\sigma_i$ is the result of a
trade-off between two opposite trends: on the one hand, the tighter the
constraint the less the solution is affected by the
corresponding rank deficiency; on the other hand, if the constraint is
too tight, the approach becomes equivalent to descoping, i.e. the
involved parameters are handled as consider parameters.

The formulation given in Eqs.~(\ref{eq_rot}) and (\ref{eq_scal}) implies
the employment of adimensional weights, which constrain the relative
accuracy of each involved parameter. To find a suitable value for the
weights, we start from a standard simulation of the relativity
experiment, obtained by estimating only 8 out of the 12 components of
the orbits of Mercury and the EMB, and we consider the ratio
between the formal accuracy of each component and the corresponding
estimated value. The results are shown in Table \ref{confr}, where we
included also the ratio of the accuracy over the estimated value of
$\mu_{\odot}$.
\begin{table}
\centering
\caption{Ratio between the formal accuracy and the estimated value for the 8 components of Mercury and the EMB orbits and for $\mu_{\odot}$. All values are normalised to $10^{-12}$.}
\label{confr} 
\begin{tabular}{ccccccccc}
\hline\noalign{\smallskip}
$\frac{\sigma(x_M)}{x_M}$ & $\frac{\sigma(y_M)}{y_M}$ & $\frac{\sigma(z_M)}{z_M}$ & $\frac{\sigma(\dot{x}_M)}{\dot{x}_M}$ & $\frac{\sigma(\dot{y}_M)}{\dot{y}_M}$ & $\frac{\sigma(\dot{z}_M)}{\dot{z}_M}$ & $\frac{\sigma(\dot{x}_E)}{\dot{x}_E}$ & $\frac{\sigma(\dot{y}_E)}{\dot{y}_E}$ & $\frac{\sigma(\mu_{\odot})}{\mu_{\odot}}$  \\
\noalign{\smallskip}\hline\noalign{\smallskip}
$0.22$ & $0.61$ & $3.1$ & $0.23$ & $0.26$ & $3.2$ & $4.4$ & $0.29$ & $0.78$ \\
\noalign{\smallskip}\hline
\end{tabular}
\end{table}
All the values range between $10^{-12}$ and $10^{-13}$, thus suitable
values to be adopted are $\sigma_i\sim 10^{-13}-10^{-14}$. In the
following, we will adopt a relative weight $\sigma_i=10^{-14}$ for
each parameter involved in the a priori constraints. Nevertheless, we
checked that adopting $\sigma_i=10^{-13}$ for each parameter, the
worsening of the global solution is negligible.

\subsection{Simulation scenario}
\label{subsec:2.2}

To perform a global simulation of the radio science experiment, we
make use of some assumptions both at simulation stage and during the
differential correction process, which are briefly described in the
following.

\paragraph{Error models.}
To simulate the observables in a realistic way, we need to make some
assumptions concerning the error sources which unavoidably affect the
observations.  We assume that the radio tracking observables are
affected only by random effects with a standard deviation of
$\sigma_{r}=15\,$cm at 300 s and
$\sigma_{\dot{r}}=1.5\times 10^{-4}\,$cm/s at 1000 s, respectively,
for Ka-band observations. The software is capable of including also a
possible systematic component to the range error model and to
calibrate for it\footnote{Two additional parameters to estimate a
  possible bias and rate over time in the range observations can be
  added to the solve-for list to avoid biasing in the solution due to
  systematic errors in ranging.}, but we did not account for this
detrimental effect in the present work, which has been partially
discussed in \cite{GiuliaMA2016}.

The accelerometer readings themselves suffer from errors of both
random and systematic origin, which can significantly bias the results
of the orbit determination. Systematic effects due to the
accelerometer readings turn out to be particularly detrimental for
the purposes of gravimetry and rotation (see, e.g., the discussion in
\cite{Cic_16}), while they induce a minor loss in accuracy for what
concerns the relativity experiment (see, e.g., \cite{G_15} and
\cite{GiuliaMA2016}). The adopted accelerometer error model, provided
by the ISA team (private communications) and the digital calibration
method applied during the differential correction process have been
extensively discussed in \cite{Cic_16}.

\paragraph{Additional rank deficiencies in the problem.}
A critical issue which significantly affects the success of the
relativity experiment concerns the high correlation between the
Eddington parameter $\beta$ and the solar oblateness
$J_{2\odot}$. Indeed, from a geometrical point of view the main
orbital effect of $\beta$ is a precession of the argument of
perihelion, which is a displacement taking place in the plane of the
orbit of Mercury, while $J_{2\odot}$ affects the precession of the
longitude of the node, producing a displacement in the plane of the
solar equator. Since the angle between the two planes is almost zero,
the two effects blend each other and the parameters turn out to be
highly correlated, causing a deterioration of the solution. A
meaningful solution to the problem is to link the PN parameters through
the Nordtvedt equation \cite{nordt}:
\begin{equation}
\eta=4(\beta-1)-(\gamma-1)-\alpha_1-\frac{2}{3}\alpha_2\,
\label{eq_N}
\end{equation}
and add such relation as an a priori constraint to the LS fit.
In such a way, the knowledge of $\beta$ is mainly determined from the
value of $\eta$ and $\gamma$, removing the correlation with
$J_{2\odot}$.  This assumption corresponds to hypothesise that gravity
is a metric theory.  

Moreover, a solar superior conjunction experiment (SCE) for the
determination of the PN parameter $\gamma$ is expected during the
cruise phase of the BepiColombo mission (see, e.g., the description in
\cite{Mil_02}), similar to the one performed by Cassini
\cite{bert}. The resulting estimate of $\gamma$ will be adopted as an
a priori constraint on the parameter in the experiment in orbit. The
complete results and a thorough discussion on the simulations of SCE 
with ORBIT14 will be presented in a future paper; however, we include
in the fit a constraint on the value of $\gamma$ given by:
$\gamma=1\pm 5\times 10^{-6}$, coming from our cruise simulations.
In this way, from Eq.~(\ref{eq_N}) it
turns out that $\beta$ is mainly determined from $\eta$, with a ratio
$1:4$ in the corresponding accuracies and a near-one correlation
between the two parameters. Indeed, this fact was already clear from
Fig.~\ref{eta_NO}: the accuracy of the two parameters shows exactly
the same behaviour as a function of the epoch of the estimate and, at
each given epoch, the ratio of the accuracies is around 4.

\paragraph{Solve-for list.}

The latest mission scenario consists of a one-year orbital phase, with
a possible extension to another year, starting from 15 March 2026. The
orbital elements of the initial Mercury-centric orbit of the MPO
orbiter are:
\begin{equation*}
1500\times 480\,\mbox{km},\,\,\,i=90^{\circ},\,\,\,\Omega=67.8^{\circ},\,\,\,\omega=16^{\circ}.
\end{equation*}
We assume that only one ground station is available for tracking, at
the Goldstone Deep Space Communications Complex in California (USA),
providing observations in the Ka-band. We solved for a total of almost
5000 parameters simultaneously in the non-linear LS fit adopting a
constrained multi-arc strategy and accounting for the
correlations. The list of solve-for parameters includes:
\begin{itemize}
\item state vector (position and velocity) of the Mercury-centric orbit
  of the spacecraft at each arc and of Mercury and Earth-Moon
  barycenter at the central epoch of the mission, in the Ecliptic Reference 
  frame at epoch J2000;
\item the PN parameters $\beta$, $\gamma$, $\eta$, $\alpha_1$,
  $\alpha_2$ and the related parameters $J_{2\odot}$, $\mu_{\odot}$,
  $\zeta$ and $GS_{\odot}$ of the Sun;
\item the calibration coefficients for the accelerometer readings at
  each arc (six parameters per arc).
\end{itemize}

\section{Numerical results}
\label{sec:3}

In this Section we describe the results of the numerical simulations
of the MORE relativity experiment. In
Sect.~\ref{subsec:3.1} we compare the two possible strategies
described in Sect.~\ref{subsec:1.1} to remove the rank deficiency of
order 4 due to the symmetry for rotation and scaling. Then, in
Sect.~\ref{subsec:3.2} we discuss the effects on the solution due to
the addition of the solar LT effect in the dynamical model, with a
particular attention on the estimate of $J_{2\odot}$.

\subsection{Removing the planetary rank deficiency}
\label{subsec:3.1}
We briefly recall the two possible strategies to remove the
approximate rank deficiency of order 4 found when we try to solve
simultaneously for the orbits of Mercury and the EMB (12 parameters)
and the solar gravitational mass $\mu_{\odot}$:
\begin{itemize}
\item strategy I (descoping)\footnote{Strategy I has been adopted until now for the MORE relativity
experiment.}: we remove 4 out of the 13 parameters from the
  solve-for list (the three position components of the EMB and the
  $z$-component of the velocity of the EMB); 
\item strategy II: we solve simultaneously for the 13 parameters by adding 4 a
  priori constraint equations in the LS fit.
\end{itemize}

In Table \ref{res_centro} the expected accuracies for the PN and
related parameters obtained following both strategies are compared
with the current knowledge of the same parameters. Table
\ref{sv_centro} provides the achievable accuracies for the state
vectors components. For all parameters, the reference date for the
estimate is the central epoch of the mission. In both tables, the
last column contains the accuracies that would be obtained if all the
state vectors components and $\mu_{\odot}$ are determined
simultaneously without any a priori constraint whatsoever. Because of
the approximate rank deficiency of order 4 (described in
Sect. \ref{subsec:1.1}), the normal matrix is still invertible, yet
the global solution is highly downgraded. 
We note indeed a loss in accuracy up to 2-3 orders of magnitude
in the components of the planetary state vectors, while an order of
magnitude is lost in the solution for $\beta$ and $\eta$. As far as the 
other relativistic parameters are concerned, it turns out that knowing
the orbits of Mercury and the EMB at the level of some meters is
sufficient to determine their value at the goal level of accuracy of
MORE.
\begin{table}[t!]
\centering
\caption{Comparison of the accuracies of PN and related parameters
  following the two possible strategies; the fourth column contains
  the current knowledge of each parameter; the last column shows the
  solution achieved by solving for all the parameters without a priori
  constraints on Mercury and EMB state vector. The accuracy of
  $\mu_{\odot}$ is in cm$^3$/s$^2$, of $\zeta$ in y$^{-1}$.}
\label{res_centro} 
\begin{tabular}{cllcl}
\hline\noalign{\smallskip}
Parameter & Strategy I & Strategy II & Current knowledge & No constraints \\
\noalign{\smallskip}\hline\noalign{\smallskip}
$\beta$ & $2.4\times 10^{-6}$ & $2.9\times 10^{-6}$ & $7\times 10^{-5}$\cite{fienga},$ \ 3.9\times 10^{-5}$\cite{park} & $2.7\times 10^{-5}$ \\
$\gamma$ & $7.6\times 10^{-7}$ & $7.6\times 10^{-7}$ & $2.3\times 10^{-5}$\cite{bert} & $ \ 7.7\times 10^{-7}$ \\
$\eta$ & $9.3\times 10^{-6}$ & $1.1\times 10^{-5}$ & $4.5\times 10^{-4}$\cite{williams} & $1.1\times 10^{-4}$ \\
$\alpha_1$  & $4.9\times 10^{-7}$  & $4.8\times 10^{-7}$ & $6.0\times 10^{-6}$\cite{iorio1} & $7.5\times 10^{-7}$ \\
$\alpha_2$  & $1.1\times 10^{-7}$  & $1.1\times 10^{-7}$ & $3.5\times 10^{-5}$\cite{iorio1} & $1.2\times 10^{-7}$ \\
$\mu_{\odot}$ & $1.0\times 10^{14}$  & $1.1\times 10^{14}$ & $8\times 10^{15}$ \cite{jpl} & $1.9\times 10^{14}$ \\
$J_{2\odot}$ & $4.9\times 10^{-9}$ & $5.0\times 10^{-9}$  & $1.2\times 10^{-8}$\cite{fienga},$\,9\times 10^{-9}$\cite{park} & $5.5\times 10^{-9}$ \\
$\zeta$ & $3.2\times 10^{-14}$ & $3.3\times 10^{-14}$ & $4.3\times 10^{-14}$ \cite{pit} & $3.5\times 10^{-14}$ \\
\noalign{\smallskip}\hline
\end{tabular}
\end{table}

\begin{table}[h!]
\centering
\caption{Comparison of the accuracies of the state vectors components
  following the two possible strategies and solving for all the
  parameters without a priori constraints. Accuracies in position are
  in cm, in velocity in cm/s.}
\label{sv_centro} 
\begin{tabular}{clll}
\hline\noalign{\smallskip}
Parameter & Strategy I & Strategy II & No constraints  \\
\noalign{\smallskip}\hline\noalign{\smallskip}
$x_M$ & 0.81 & 0.65 & $4.50\times 10^2$ \\
$y_M$ & 3.6 & 3.0 & $2.68\times 10^2$ \\
$z_M$ & 4.2 & 2.2 & $1.633\times 10^3$ \\
$x_E$  & -- & 0.70 & $5.89\times 10^1$ \\
$y_E$  & -- & 2.8  & $1.093\times 10^3$ \\
$z_E$ & --  & 4.3 & $4.275\times 10^3$ \\
$\dot{x}_M$ & $7.3\times 10^{-7}$  & $3.6\times 10^{-7}$  & $2.83\times 10^{-4}$ \\
$\dot{y}_M$ & $6.1\times 10^{-7}$  & $2.4\times 10^{-7}$ & $2.58\times 10^{-4}$ \\
$\dot{z}_M$ & $1.5\times 10^{-6}$ & $8.1\times 10^{-7}$ & $1.01\times 10^{-3}$ \\
$\dot{x}_E$ & $5.0\times 10^{-7}$ & $1.2\times 10^{-7}$ & $2.16\times 10^{-4}$ \\
$\dot{y}_E$ & $8.6\times 10^{-7}$  & $9.2\times 10^{-7}$ & $8.26\times 10^{-6}$ \\
$\dot{z}_E$ & --  & $7.3\times 10^{-7}$ & $6.27\times 10^{-4}$ \\
\noalign{\smallskip}\hline
\end{tabular}
\end{table}

\begin{figure}[t]
  \includegraphics[width=0.50\textwidth]{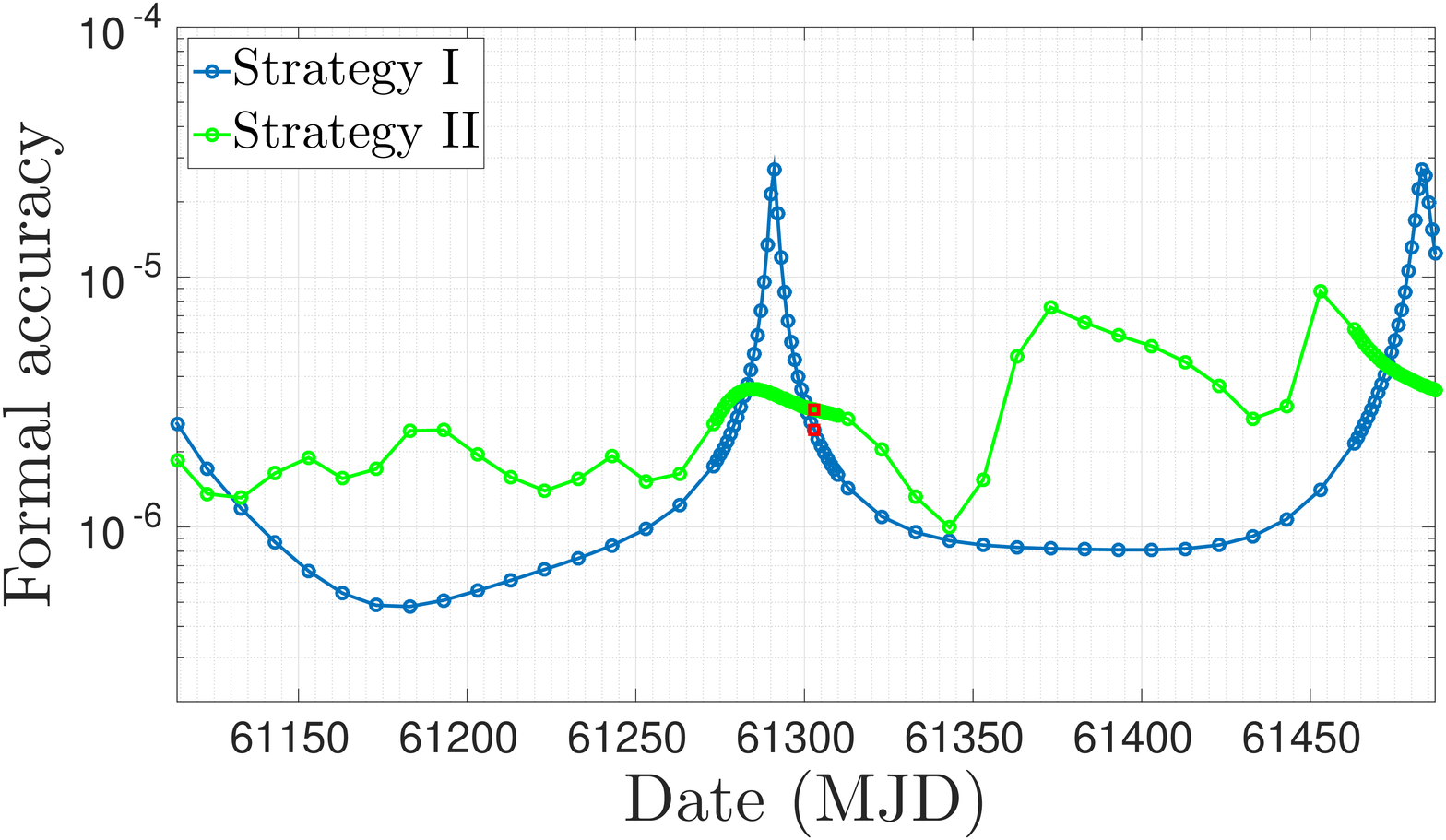}
  \includegraphics[width=0.50\textwidth]{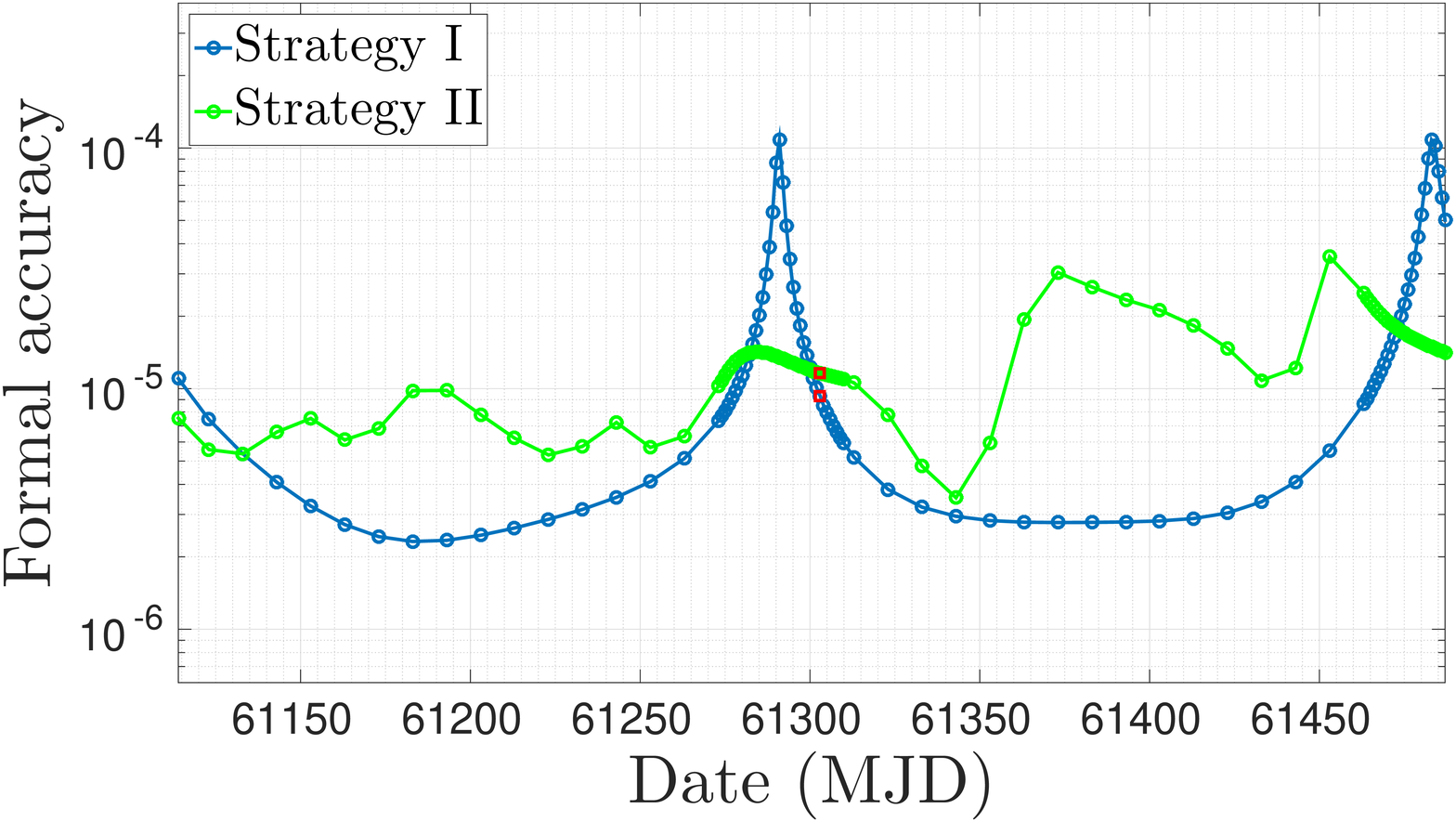}
  \caption{Comparison of the formal accuracy of $\beta$ (left) and
    $\eta$ (right) as a function of the epoch of the estimate (in MJD)
    over the mission time span adopting strategy I (blue curve) and
    strategy II (green curve). In red the value of the accuracy for
    the estimate at central epoch.}
  \label{eta_SI}
\end{figure}

The results achievable with strategies I and II are
almost comparable and represent a significant improvement with respect
to the current knowledge (see the discussion in \cite{universe} for a
comparison with the actual knowledge). This is true if orbit
determination is performed at the central epoch of the orbital
mission. Indeed, from Fig. \ref{eta_NO} we have seen that, adopting
strategy I, there is a strong dependency of the solution from the
epoch. In Fig.  \ref{eta_SI} we compare the behaviour of the formal
accuracy of $\beta$ (on the left) and $\eta$ (on the right) adopting
strategy I (blue curve) and strategy II (green curve). The red circle
refers to the estimate at the central epoch (MJD 61303).
Choosing the second strategy, we observe that the dependency of the
accuracy from the epoch of the estimate is definitely reduced. If we
consider the evolution of the formal of $\eta$ from the beginning of
the orbital mission up to MJD 61350, the variability in case of
strategy I spans from a minimum of $\sigma(\eta)=2.3\times 10^{-6}$ to
a maximum of $\sigma(\eta)=1.1\times 10^{-4}$, while adopting strategy
II the formal accuracy ranges from a minimum of
$\sigma(\eta)=3.5\times 10^{-6}$ to a maximum of
$\sigma(\eta)=1.4\times 10^{-5}$, with a net variability of only a
factor 4 instead of a factor 50. If the orbits of Mercury and the EMB
are determined in the second part of the mission, we observe a
stronger variability in the accuracies adopting the second
approach. Such behaviour could suggest that some degeneracy is still
affecting the orbit determination problem. This issue will be
investigated in the future. Nevertheless, the standard strategy of orbit
determination codes is to adopt, as the reference epoch, the initial
or the central date, thus for the purpose of our simulations we can
ignore the behaviour of the curves in the second half of the mission
time span.

Of course, if the mission scenario is exactly the one adopted in our
simulations, i.e. assuming the beginning of scientific operations on
15 March 2026 and the end on 21 March 2027 (corresponding to 365
observed arcs\footnote{For the definition of observed arc see, e.g.,
  \cite{Cic_16}.}), choosing strategy I or II does not lead to
significant differences in the solution. However, Fig. \ref{eta_SI}
states that strategy II provides a more stable solution. Indeed, as an
example, in Table \ref{2weeks} we show the results for the accuracy of
relativity parameters in the hypothesis of moving up the beginning of
the orbital experiment by approximately two weeks, on 3 March 2026,
still keeping the one-year duration.
\begin{table}[t!]
\centering
\caption{Comparison of the accuracies of PN and related parameters
  following the two possible strategies in the hypothesis of beginning
  of scientific operations in orbit on 3 March 2026 instead of 15
  March 2026. The last two columns show the ratio of accuracy
  attained, for each parameter, on the scenario of 3 March over that
  achieved on the scenario of 15 March, in the case of strategy I and
  II, respectively. The accuracy of $\mu_{\odot}$ is in cm$^3$/s$^2$,
  of $\zeta$ in y$^{-1}$.}
\label{2weeks} 
\begin{tabular}{cllll}
\hline\noalign{\smallskip}
Parameter & Strategy I & Strategy II & Ratio I & Ratio II \\
\noalign{\smallskip}\hline\noalign{\smallskip}
$\beta$ & $3.0\times 10^{-5}$ & $2.3\times 10^{-6}$ & $12.5$ & $0.79$ \\
$\gamma$ & $6.4\times 10^{-7}$ & $7.7\times 10^{-7}$ & $0.84$ & $1.0$ \\
$\eta$ & $1.2\times 10^{-4}$ & $8.7\times 10^{-6}$ & $12.5$ & $0.79$ \\
$\alpha_1$  & $7.6\times 10^{-7}$  & $4.4\times 10^{-7}$  & $1.5$ & $0.92$ \\
$\alpha_2$  & $9.6\times 10^{-8}$  & $8.0\times 10^{-8}$ & $0.87$ & $0.73$ \\
$\mu_{\odot}$ & $1.8\times 10^{14}$  & $7.9\times 10^{13}$ & $1.8$ & $0.72$ \\
$J_{2\odot}$ & $4.7\times 10^{-9}$ & $4.3\times 10^{-9}$  & $0.9$ & $0.86$ \\
$\zeta$ & $2.6\times 10^{-14}$ & $2.7\times 10^{-14}$ & $0.81$ & $0.82$ \\
\noalign{\smallskip}\hline
\end{tabular}
\end{table}
The last two columns of Table \ref{2weeks} shows the ratio between the
accuracy achieved for each parameter in the scenario of 3 March 2026
and the one of the 15 March 2026 scenario, for strategies I and II,
respectively. It is clear that adopting the second approach, a slight
variation in the initial date of the mission in orbit leads only to
slight variations in the accuracy of the relativity parameters, as it
has to be. Conversely, in the case of the first strategy the solution
turns out to be less stable. Indeed, the accuracy of $\beta$ and $\eta$
varies by an order of magnitude between the two scenarios, weakening
the reliability of the achieved results.

For completeness, Table \ref{corr} shows the correlations between PN
and related parameters in the case of strategy I (top) and strategy II
(bottom). Values higher than 0.8 have been highlighted.
\begin{table}[h!]
\caption{Correlations between PN and related parameters in the case of
  strategy I (top) and strategy II (bottom). Values higher than 0.8
  have been highlighted.}
\label{corr} 
\begin{tabular}{lcccccccc}
\hline\noalign{\smallskip}
 & $\beta$ & $\gamma$ & $\eta$ & $\alpha_1$ & $\alpha_2$ & $\mu_{\odot}$ & $J_{2\odot}$ & $\zeta$ \\
\noalign{\smallskip}\hline\noalign{\smallskip}
$\zeta$ & $<0.1$ & $0.12$ & $<0.1$ & $0.12$ & $0.49$ & $0.74$ & $0.76$ & -- \\
$J_{2\odot}$ & $<0.1$ & $< 0.1$ & $<0.1$ & $<0.1$ & $0.26$ & $\mathbf{0.86}$ & --  \\
$\mu_{\odot}$ & $0.22$ & $< 0.1$ & $0.22$ & $0.42$ & $0.38$ & -- \\
$\alpha_2$  & $0.44$ & $0.14$ & $0.46$ & $0.28$ & --  \\
$\alpha_1$  & $0.25$ & $0.12$ & $0.21$ & -- \\
$\eta$ & $\mathbf{0.99}$ & $0.56$ & -- \\
$\gamma$ & $0.62$ & -- \\
$\beta$ & -- \\
\noalign{\smallskip}\hline\noalign{\smallskip}
 & $\beta$ & $\gamma$ & $\eta$ & $\alpha_1$ & $\alpha_2$ & $\mu_{\odot}$ & $J_{2\odot}$ & $\zeta$ \\
\noalign{\smallskip}\hline\noalign{\smallskip}
$\zeta$ & $0.63$ & $0.11$ & $0.64$ & $< 0.1$ & $0.51$ & $0.76$ & $0.77$ & -- \\
$J_{2\odot}$ & $0.54$ & $< 0.1$ & $0.55$ & $0.11$ & $0.29$ & $\mathbf{0.86}$ & --  \\
$\mu_{\odot}$ & $0.76$ & $< 0.1$ & $0.76$ & $0.35$ & $0.42$ & -- \\
$\alpha_2$  & $0.73$ & $0.13$ & $0.73$ & $0.24$ & --  \\
$\alpha_1$  & $0.36$ & $0.17$ & $0.31$ & -- \\
$\eta$ & $\mathbf{0.99}$ & $< 0.1$ & -- \\
$\gamma$ & $0.16$ & -- \\
$\beta$ & -- \\
\noalign{\smallskip}\hline
\end{tabular}
\end{table}
In both cases we find a high correlation only between the two physical
parameters of the Sun, i.e. $\mu_{\odot}$ and $J_{2\odot}$, and
between $\beta$ and $\eta$, whose correlation is near 1 due to the
assumption that PN parameters are linked through the Nordtvedt
equation.  In general, correlations between the parameters, although
restrained, are higher in the case of strategy II. This fact was
expected since, from Table \ref{res_centro}, formal accuracies at the
central epoch are slightly worse than adopting strategy
I. Nevertheless, except the correlation between
$\mu_\odot$-$J_{2\odot}$ and $\beta$-$\eta$, they are always lower
than 0.8.

\subsection{Solar LT effect and the determination of $J_{2\odot}$}
\label{subsec:3.2}

In Sect.~\ref{subsec:1.2} we showed that the solar LT effect on
Mercury produces a signal with a peak-to-peak amplitude up to about
ten meters after one year, hence it should be taken into account in
the BepiColombo radio science data processing, otherwise it would
alias the recovery of other effects, as already pointed out in
\cite{iorio2}. In that paper it was also underlined that the
measurement of the solar quadrupole $J_{2\odot}$ at the $1\%$ level or
better, which is one of the goals of MORE, cannot be performed aside
from accounting for the solar LT effect; the impact of
neglecting the gravitomagnetic field of the Sun may affect indeed the
determination of $J_{2\odot}$ at the $12\%$ level. Moreover, in
\cite{park} the authors observe that, processing three years of
ranging data to MESSENGER by explicitly modelling the gravitomagnetic
field of the Sun, the small precession of the perihelion of Mercury
induced by solar LT turns out to be highly correlated with the
precession due to $J_{2\odot}$.  

In this section we investigate two different aspects of the problem
with BepiColombo MORE: firstly, we measure the impact on the estimated
value of $J_{2\odot}$ if we do not include the solar LT in the
dynamical model; secondly, we check whether solving for $GS_{\odot}$
introduces some weakness in the orbit determination problem, for
instance deteriorating the formal uncertainties of the other
parameters, especially $J_{2\odot}$.

In order to address the first matter, we simulated one year of
BepiColombo observations including the solar LT effect and then we
applied the differential corrections in two different cases: (i) we
included solar LT in the corrector model; (ii) we did not include
solar LT in differential corrections. The set of estimated parameters
is the same of Sect~\ref{subsec:2.2}, except for the solar angular
momentum, which is assumed at the nominal value $S_{\odot}=1.92\times
10^{48}\,$g$\,$cm$^2$/s \cite{iorio4}. The results for the estimated
value and formal accuracy of $J_{2\odot}$ in the two cases are shown
in Table \ref{J2_1}.
\begin{table}
\centering
\caption{Estimated value and formal accuracy of $J_{2\odot}$ with
  solar LT on and off, respectively, in differential correction
  stage. The parameter $GS_{\odot}$ is not determined and it is
  assumed at its nominal value.}
\label{J2_1} 
\begin{tabular}{lll}
\hline\noalign{\smallskip}
Case & Estimated value & $\sigma(J_{2\odot})$  \\
\noalign{\smallskip}\hline\noalign{\smallskip}
LT ON & $1.992\times 10^{-7}$ & $6.0\times 10^{-10}$ \\
LT OFF & $1.837\times 10^{-7}$ & $6.0\times 10^{-10}$ \\
\noalign{\smallskip}\hline
\end{tabular}
\end{table}
As expected, the formal accuracy is the same in both cases, while the
estimated value of $J_{2\odot}$ at convergence is
different\footnote{The nominal value of $J_{2\odot}$ in simulation has
  been set to $2.0\times 10^{-7}$.}. More precisely, we observe that
neglecting the solar LT effect on the orbit of Mercury (second
simulation) introduces a bias in the estimated value of $J_{2\odot}$
as large as $27\sigma$. The effect on the other parameters is only
marginally relevant: we remarked a bias in $\mu_{\odot}$ of $\sim
5\sigma$ and in some components of the orbit of Mercury, of the same
amount.  On the contrary, in the first simulation the estimated value
of $J_{2\odot}$ lies within $1.3\sigma$ with respect to the nominal
value. In conclusion, this test confirms that the solar LT
acceleration produces effects on the orbit of Mercury which can be
absorbed by $J_{2\odot}$, if not properly modelled. Under no
circumstances should the LT effect be neglected for the BepiColombo
MORE experiment.

Now that we proved that it is crucial to include the gravitomagnetic
acceleration due to the Sun in the dynamical model, we go on to
discuss the second point.  We introduce the Lense-Thirring parameter
$GS_{\odot}$ in the solve-for list: due to the high correlation with
$J_{2\odot}$, we expect to find a significant worsening in the
solution for the solar oblateness. A similar behaviour was already
found in the case of the mission Juno and described in \cite{serra}.
We considered three explanatory cases: (i) $J_{2\odot}$ and
$GS_{\odot}$ are determined simultaneously without any a priori
information on their values (same setup of Sect.~\ref{subsec:2.2});
(ii) the value of $J_{2\odot}$ is a priori constrained to its present
knowledge $2\pm 0.12\times 10^{-7}$ (cf. \cite{fienga}); (iii) the
value of $GS_{\odot}$ is a priori constrained to $10\%$
level\footnote{From heliosismology, the angular momentum of the Sun
  can be constrained significantly better than the 10\% level (see,
  e.g., \cite{pijpers}), thus our assumption is fully acceptable and
  is consistent with what done in \cite{park}.}.
\begin{table}
\centering
\caption{Achievable accuracy on $J_{2\odot}$ and correlation with
  $GS_{\odot}$ in the following cases: (i) $J_{2\odot}$ and
  $GS_{\odot}$ are determined assuming no a priori information on
  their value; (ii) $J_{2\odot}$ is constrained to the a priori value
    $2\pm0.12\times 10^{-7}$ (present knowledge); (iii) $GS_\odot$ is
    a priori constrained to $10\%$ of its value.}
\label{GS} 
\begin{tabular}{lccc}
\hline\noalign{\smallskip}
 & Case (i) & Case (ii) & Case (iii)  \\
\noalign{\smallskip}\hline\noalign{\smallskip}
$\sigma(J_{2\odot})$ & $5.0\times 10^{-9}$ & $4.6\times 10^{-9}$ & $1.7\times 10^{-9}$ \\
correlation with $GS_{\odot}$ & 0.9928 & 0.9919 & 0.9354 \\
\noalign{\smallskip}\hline
\end{tabular}
\end{table}
The results are shown in Table \ref{GS}.
The simultaneous determination of $J_{2\odot}$ and $GS_\odot$ without
any a priori (case (i)) leads to a 0.99 correlation between the two
parameters, as expected. As a result, the solution with respect to the
first row of Table \ref{J2_1} is downgraded by almost an order of
magnitude. Add an a priori on $J_{2\odot}$ at the level of the
current knowledge (case (ii)) does not change much the result, 
as the correlation between $J_{2\odot}$ and $GS_{\odot}$
does not decrease significantly. Conversely, a rather weak constraint on
$GS_{\odot}$ (case (iii)) is capable of significantly improving the
solution, breaking the correlation between the two parameters (from 0.99 to 0.93). 
A tighter constraint on $GS_{\odot}$ would provide a further improvement of the
results. As a conclusion, we can state that the achievable accuracy on
$J_{2\odot}$ will be mainly limited by the knowledge of the solar
angular momentum.

\section{Conclusions and remarks}
\label{concl}
The present paper addresses two critical aspects of the BepiColombo
relativity experiment we aim to solve. The first one concerns the
approximate rank deficiency of order 4 found in the Earth and Mercury
orbit determination problem. In particular, we highlighted that,
according on how the rank deficiency is cured, the dependency of the
PN parameters $\beta$ and $\eta$ from the epoch of the estimate can be
highly pronounced. As a consequence, the reliability of the solution
can be compromised. We considered two possible strategies: the set of
13 critical parameters (initial conditions of Mercury and EMB and the
gravitational mass of the Sun) can be reduced to only 9 parameters to
be determined, as done up to now in the relativity experiment
settings, or we can solve for the whole set of parameters providing 4
a priori constraint equations in input to the differential correction
process. We concluded that, although by chance the present mission
scenario does not imply considerable differences between the two
strategies, the second strategy leads to a more stable solution and,
thus, is the more advisable approach.

Secondly, we studied the impact on the determination of the solar oblateness
parameter $J_{2\odot}$ of a failure to include the solar LT perturbation in 
Mercury's dynamical model. The parameter $J_{2\odot}$ turns out to be 
highly correlated with the LT parameter $GS_{\odot}$, containing the solar 
angular momentum. We pointed out that neglecting the solar
LT effect leads to a considerable bias in the estimated value of $J_{2\odot}$, 
and to an illusory high accuracy in the determination of the same parameter. 
Nevertheless, we have shown that including in the LS fit some reasonable a priori
information on $GS_{\odot}$ can help contain the deterioration of the solution for
$J_{2\odot}$.

  The results of the research presented in this paper have been
  performed within the scope of the Addendum n. I/080/09/1 of the
  contract n. I/080/09/0 with the Italian Space Agency.




\end{document}